\documentclass[aps,twocolumn,showpacs,showkeys]{revtex4}
\usepackage{graphicx}

\begin{document}

\title{The Feigenbaum's $\delta$ for a high dissipative bouncing ball
model}

\author{Diego F.\ M.\ Oliveira}
\email{dfmo@rc.unesp.br}
\author{Edson D.\ Leonel}
\affiliation{Departamento de Estat\'istica, Matem\'atica Aplicada e
Computa\c c\~ao, Universidade Estadual Paulista, CEP, 13506-900, Rio
Claro, SP, Brazil}

\date{\today}\widetext



\begin{abstract}
We have studied a dissipative version of a one-dimensional Fermi
accelerator model. The dynamics of the model is described in terms
of a two-dimensional, nonlinear area-contracting map. The dissipation
is introduced via innelastic collisions of the particle with the
walls and we consider the dynamics in the regime of high dissipation.
For such a regime, the model exhibits a route to chaos known as period
doubling and we obtain a constant along the bifurcations so called the
Feigenbaum's number $\delta$.
\end{abstract}
\maketitle

\section{Introduction}

The origin of cosmic radiation and a mechanism trough which it could
acquire enormous energies has intrigued many physicists and
mathematicians along the last decades. In particular, in the year of
1949 Enrico Fermi \cite{ref1} proposed a very simple model that
qualitatively describes a process in which charged particles were {\it
bounced} via interaction with moving magnetic fields. This heuristic
idea was then later modified to encompass in a suitable model that
could give quantitative results on the original Fermi's model. After
that, many different versions of the model were proposed
\cite{ref2,ref3,ref4,ref5,ref6,ref7,NEW1,NEW2} and studied in different
context and considering many approaches and modifications
\cite{ref8,ref9,ref10,ref11,ref12,ref13, ref14,ref15}. One of them is
the well known Fermi-Ulam model (FUM). Such model consists of a
classical point-like particle moving between two walls in the
total absence of any external field. One of the walls is
considered to be fixed while the other one moves periodically
in time. Despite this simplicity, the structure of the phase space is
rather complex and it includes a large chaotic sea surrounding KAM
islands and a set of invariant spanning curves limiting the energy of a
bouncing particle. However, the introduction of innelastic collisions
on this model is enough to destroy such a mixed structure and the
system exhibits attractors. Depending on the initial conditions and
control parameters, one can observe a chaotic attractor characterised
by a positive Lyapunov exponent. By a suitable control parameter
variation, the chaotic attractor might be destroyed via a crisis event
\cite{ref19,ref20,add1}. After the destruction, the chaotic attractor
is replaced by a chaotic transient \cite{ref20}.

In this paper, we revisit a dissipative version of the Fermi-Ulam
model seeking to understand and describe the behaviour of the dynamics
for the regime of high dissipation. Moreover, we investigate a
phenomenon quite common observed for a large variety of dissipative and
nonlinear systems known as {\it period doubling route to chaos}. This
route shows a sequence of doubling bifurcations connecting
regular-periodic to chaotic behaviour. The main goal of this paper is
then to examine and show that the Fermi-Ulam model obey the same
convergence ratio as that obtained by Feigenbaum \cite{ref21}
for the dynamics of the logistic map. Such a constant value has been
further called as ``The Feigenbaum's $\delta$''.

The organisation of this paper is as follows: In section \ref{sec2} we
present all the details needed for the construction of the nonlinear
mapping. Our numerical results are discussed in section \ref{sec3} and
our conclusions and final remarks are drawn in section \ref{sec4}.

\section{A dissipative Fermi-Ulam model}
\label{sec2}

Let us describe the model and obtain the equations of the mapping. The
model we are dealing with consists of a classical point-like particle
confined in and bouncing betwen two rigid walls. One of them is assumed
to be fixed at the position $x=l$ and the other one moves periodically
in time according to $x_w(t)=\varepsilon\cos(\omega t)$. Thus, the
moving wall velocity is given by $v_w(t)=-\varepsilon\omega
\sin(\omega t)$ where $\varepsilon$ denotes the amplitude of
oscillation and $\omega$ is the angular frequency of the moving wall,
respectively. Aditionally, the motion of the particle does not suffer
influence of any external field. We assume that all the collisions with
both walls are inelastic. Thus, we introduce a restitution coefficient
for the fixed wall as $\alpha\in[0,1]$ while for the moving wall we
consider $\beta\in[0,1]$. The completely inelastic collision happens
when $\alpha=\beta=0$ and that a single collision is enough to
terminate all the particle's dynamics. On the other hand, when
$\alpha=\beta=1$, corresponding to a complete elastic collision, all
the results for the nondissipative case are recovered \cite{ref15}.

We describe the dynamics of the particle using a map $T$ where the
dynamical variables are $(v_n,t_n)$ and $v_n$ is the velocity of the
particle at the instant $t_n$. The index $n$ denotes the $n^{th}$
collision of the particle with the moving wall. Starting then with an
initial condition $(v_n,t_n)$, with initial position of the particles
given by $x_p(t)=\varepsilon \cos(\omega t_n)$ with $v_n>0$, the
dynamics is evolved and the map $T$ gives a new pair of
$(v_{n+1},t_{n+1})$ at the $(n+1)^{th}$ collision. It is important to
say that we have three control parameters, namely: $\varepsilon$, $l$
and $\omega$ and that the dynamics does not depend on all of them.
Thus it is convenient to define dimensionless and more appropriated
variables. Therefore, we define $\epsilon=\varepsilon/l$,
$V_n=v_n/{\omega l}$ and measure the time in terms of the number of
oscillations of the moving wall, consequently $\phi_n=\omega t_n$.
Incorporating this set of new variables into the model, the map $T$ is
written as
\begin{equation}
T:\left\{\begin{array}{ll}
V_{n+1}=V_n^*-(1+\beta)\epsilon\sin(\phi_{n+1})~~\\
\phi_{n+1}=\phi_n+\Delta T_n~~{\rm mod (2\pi)}\\
\end{array}
\right.~,
\label{eq1}
\end{equation}
where the expressions for both $V_n^*$ and $\Delta T_n$ depend on what
kind of collision occurs. There are two different possible situations,
namely: (i) multiple collisions with the moving wall and, (ii) single
collision with the moving wall. Considering the first case, the
expressions are $V_n^*=-\beta V_n$ and $\Delta T_n=\phi_c$, where
$\phi_c$ is obtained as the smallest solution of $G(\phi_c)=0$ with
$\phi_c\in(0,2\pi]$. A solution for $G(\phi_c)=0$ is equivalent to
have that the position of the particle is the same as that of the
moving wall at the instant of the impact. The function $G(\phi_c)$ is
given by
\begin{equation} 
G(\phi_c)=\epsilon\cos(\phi_n+\phi_c)-\epsilon\cos(\phi_n)-V_n\phi_c~.
\label{g}
\end{equation}
If the function $G(\phi_c)$ does not have a root in the interval
$\phi_c\in(0,2\pi]$, we can conclude that the particle leaves the
collision zone (the collision zone is defined as the interval
$x\in[-\epsilon,\epsilon]$) without suffering a successive collision.

Considering now the case (ii), i.e. the case where the particle leaves
the collision zone, the corresponding expressions are
$V_n^*=-\alpha\beta V_n$ and $\Delta T_n=\phi_c + \phi_r+\phi_l$, with
the auxiliary terms given by
\begin{equation}
\phi_r={{1-\epsilon\cos(\phi_n)}\over{V_n}}~~,~~
\phi_l={{1-\epsilon}\over{\alpha V_n}}~.
\label{aux}
\end{equation}
Finally, the term $\phi_c$ is obtained as the smallest solution of
$F(\phi_c)=0$ for $\phi_c\in[0,2\pi)$. The expression of $F(\phi_c)$ is
given by
\begin{equation}
F(\phi_c)=\epsilon\cos(\phi_n+\phi_r+\phi_l+\phi_c)-\epsilon+\alpha
V_n\phi_c~.
\label{f}
\end{equation}
Both, Eq. (\ref{g}) and Eq. (\ref{f}) must be solved numerically.

After some algebra it is easy to shown that the determinant of the
Jacobian Matrix for the case (i) is given by
\begin{equation}
det(J)=\beta^2\left[{{V_n+\epsilon\sin(\phi_n)}
\over{V_{n+1}+\epsilon\sin(\phi_{n+1})}}\right]~,
\label{det_s}
\end{equation}
while for the case (ii) we obtain
\begin{equation}
det(J)=\alpha^2\beta^2\left[{{V_n+\epsilon\sin(\phi_n)}
\over{V_{n+1}+\epsilon\sin(\phi_{n+1})}}\right]~.
\label{det_ns}
\end{equation}
We therefore can conclude based on the above result that area
preservation is obtained only for the case of $\alpha=\beta=1$.

\section{Numerical Results}
\label{sec3}

\begin{figure}[b]
\centerline{\includegraphics[width=1.0\linewidth]{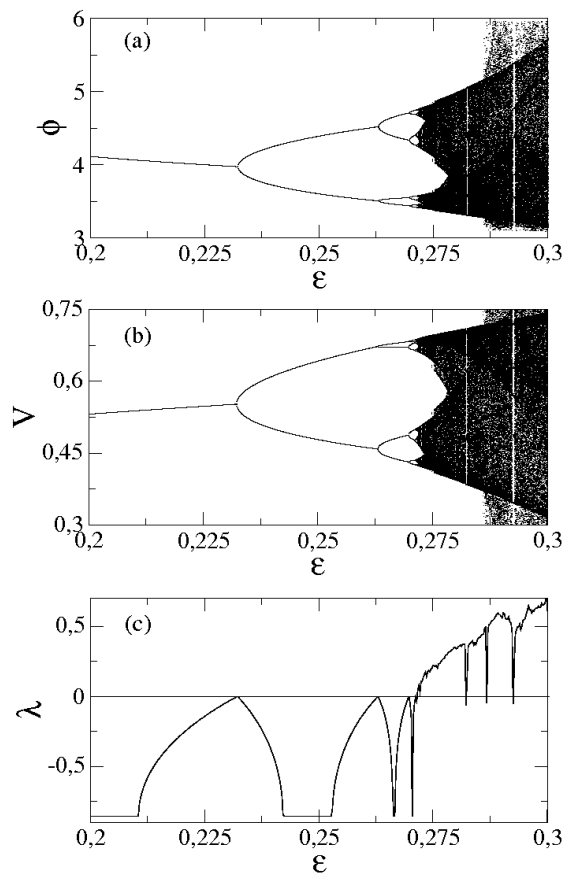}}
\caption{Bifurcation cascade for (a) $\phi$ and (b) $V$ both plotted
against $\epsilon$. In (c) it is shown the Lyapunov exponent associated
to (a) and (b). The damping coefficients used for the construction of
the figures (a), (b) and (c) were $\alpha=0.85$ and $\beta=0.50$.}
\label{fig1}
\end{figure}

In this section we discuss some numerical results considering the case
of high dissipation. We call as high dissipation the situation in
which a particle loses more than $50\%$ of its energy upon colision
with the moving wall. Moreover, we have considered as fixed the values
of the damping coefficients $\alpha=0.85$ and $\beta=0.50$. To
illustrate that the dynamics of the system exhibits doubling
bifurcation cascade, it is shown in Fig. \ref{fig1}(a)
the behaviour of the asymptotic velocity plotted against the control
parameter $\epsilon$, where the sequence of bifurcations is evident. A
similar sequence is also observed for the asymptotic variable $\phi$,
as it is shown in Fig. \ref{fig1}(b). Note that all the bifurcations
of same period in (a) and (b) happen for the same value of the
parameter $\epsilon$. At the point of bifurcations, one can also
observe that the positive Lyapunov exponent shows null value, as it is
shown in Fig. \ref{fig1}(c). It is well known that the Lyapunov
exponents are important tool that can be used to classify orbits as
chaotic. As discussed in \cite{ref22}, the Lyapunov exponents are
defined as
\begin{equation}
\lambda_j=\lim_{n\rightarrow\infty}{1\over{n}}\ln|\Lambda_j|~~,~~j=1,
2~~,
\label{eq4}
\end{equation}
where $\Lambda_j$ are the eigenvalues of
$M=\prod_{i=1}^nJ_i(V_i,\phi_i)$ and $J_i$ is the Jacobian matrix
evaluated over the orbit $(V_i,\phi_i)$. However, a direct
implementation of a computational algorithm to evaluate Eq. (\ref{eq4})
has a severe limitation to obtain $M$. Even in the limit of short $n$,
the components of $M$ can assume very different orders of magnitude for
chaotic orbits and periodic attractors, yielding impracticable the
implementation of the algorithm. In order to avoid such problem, we
note that $J$ can be written as $J=\Theta T$ where $\Theta$ is an
ortoghonal matrix and $T$ is a right up triangular matrix. Thus we
rewrite $M$ as $M=J_nJ_{n-1}\ldots J_2\Theta_1\Theta_1^{-1}J_1$, where
$T_1=\Theta_1^{-1}J_1$. A product of $J_2\Theta_1$ defines a new
$J_2^{\prime}$. In a next step, it is easy to show that
$M=J_nJ_{n-1}\ldots J_3\Theta_2\Theta_2^{-1}J_2^{\prime}T_1$. The same
procedure can be used to obtain $T_2=\Theta_2^{-1}J_2^{\prime}$ and so
on. Using this procedure, the problem is reduced to evaluate the
diagonal elements of $T_i:T_{11}^i,T_{22}^i$. Finally, the Lyapunov
exponents are given by
\begin{equation}
\lambda_j=\lim_{n\rightarrow\infty}{1\over{n}}\sum_{i=1}^n
\ln|T_{jj}^i|~~,~~j=1,2~~.
\label{eq5}
\end{equation}

If at least one of the $\lambda_j$ is positive then the orbit is
classified as chaotic. We can see in Fig. (\ref{fig1}) (c) the behavior
of the Lyapunov exponents corresponding to both Figs.
(\ref{fig1})(a,b). It is also easy to see that when the bifurcations
happen, the exponent $\lambda$ assumes the zero value at same values of
the control parameter $\epsilon$ where the bifurcation hold. Another
interesting observation is that the Lyapunov exponent, in some regions,
assumes a constant and negative values, such behavior occurs because
the eigenvalues of the Jacobian Matrix become complex numbers
(imaginary numbers)).

Let us now use the points where the Lyapunov exponent assumes the null
value to characterise a very interesting property of the model. As it
was shown by Feigenbaum \cite{ref21}, along the bifurcations the
dynamics exhibits an universal feature. It implies that all the
bifurcations happen at the same rate of convergence for the bifurcation
diagram changing from periodic to chaotic behavior. The procedure used
to obtain the Feigenbaum value $\delta$ consists of: let $\epsilon_1$
represents the control parameter value at which period-1 gives birth to
a period-2 orbit, $\epsilon_2$ is the value where period-2 changes to
period-4 and so on. In general the parameter $\epsilon_n$ corresponds
to the control parameter value at which a period-$2^n$ orbit is born.
Thus, we write the Feigenbaum's $\delta$ as
\begin{equation}
\delta_{n}={\lim_{n\gg1}{{\epsilon_{n}-\epsilon_{n-1}} \over
{\epsilon_{n+1}-\epsilon_{n}}}}.
\label{fei}
\end{equation}

The numerical value for the constant number $\delta$
obtained by Eq. (\ref{fei}) and considering a sufficient large values
for $n$ is $\delta=4.66920161\ldots$. Considering the numerical data
obtained through the Lyapunov exponents calculation, the Feigenbaum's
$\delta$ obtained for the Fermi-Ulam model is $\delta=4.64(4)$. We have
considered in our simulations only the bifurcations of fourth to eighth
order since the numerical results are very hard to be obtained for
higher orders in the bifurcations. Our result is in a good agreement
with the Feigenbaum's universal $\delta$.

\section{Conclusion}
\label{sec4}

As a summary, we have studied a dissipative version of the Fermi-Ulam
model. We introduce dissipation into the model through inelastic
collisions with both walls. We have shown that for regimes of high
dissipation, the model exhibits a sequence of doubling bifurcation
cascade. For this cascade, we obtained the so called Feigenbaum's
number as $\delta=4.64(4)$.

\section*{Acknowledgements}
DFMO gratefully acknowledges Conselho Nacional de Desenvolvimento
Cient\'{\i}fico e Tecnol\'ogico -- CNPq. EDL is grateful to FAPESP, 
CNPq and FUNDUNESP, Brazilian agencies.


\begin{thebibliography}{99}

\bibitem{ref1} E. Fermi, Phys. Rev. {\bf{75}} 1169 (1949).

\bibitem{ref2} Ulam S 1961 Proc. 4th Berkeley Symposium on Math,
Statistics and Probability vol 3 (Berkeley, CA: California University
Press) p 315.

\bibitem{ref3} A. J. Lichtenberg, M.A. Lieberman, ``Regular
and Chaotic Dynamics'' (Appl. Math. Sci. {\bf{38}}, Springer Verlag,
New York, (1992).

\bibitem{ref4} A. J. Lichtenberg, M.A. Lieberman and R. H. Cohen,
Physica D {\bf{1}}, 291 (1980).

\bibitem{ref5} P. J. Holmes, J. Sound and Vibr. {\bf{84}}, 173 (1982).

\bibitem{ref6} K. Y. Tsang and M. A. Lieberman, Physica D {\bf{11}},
147 (1984).

\bibitem{ref7} M. A. Lieberman and K. Y. Tsang, Phys. Rev. Lett.
{\bf{55}}, 908 (1985).

\bibitem{NEW1} J. K. L. da Silva, D. G. Ladeira, E. D. Leonel,
P. V. E. McClintock, S. O. Kamphorst, Brazilian Journal of Physics
{\bf{36}}, 700 (2006).

\bibitem{NEW2} E. D. Leonel, D. F. M. Oliveira, R. E. de Carvalho,
Physica A {\bf{386}}, 73 (2007).

\bibitem{ref8} R. M. Everson, Physica D {\bf{19}}, 355 (1986).

\bibitem{ref9} J. V. Jos\'e and R. Cordery, Phys. Rev. Lett. {\bf{56}},
290 (1986).

\bibitem{ref10} G. A. Luna-Acosta, Phys. Rev. {\bf{42}}, 7155 (1990).

\bibitem{ref11} S. T. Dembinski, A. J. Makowski and P. Peplowski, Phys.
Rev. Lett. {\bf{70}}, 1093 (1993).

\bibitem{ref12} E. D. Leonel, P. V. E. McClintock and J. K. L. da
Silva, Phys. Rev. Lett. {\bf{93}}, 014101  (2004).

\bibitem{ref13} E. D. Leonel and P. V. E. McClintock, J. Phys. A
{\bf{38}}, 823 (2005).

\bibitem{ref14} D. G. Ladeira, J. K. L. da Silva, Phys. Rev. E
{\bf{73}}, 026201 (2006).

\bibitem{ref15} E. D. Leonel, J. K. L. da Silva and S. O. Kamphorst,
Physica A {\bf{331}}, 435 (2004).

\bibitem{ref19} E. D. Leonel and P. V. E. McClintock, J. Phys. A,
{\bf{38}}, L425 (2005).

\bibitem{ref20} E. D. Leonel and R. Egydio de Carvalho, Phys. Lett. A,
{\bf{364}}, 475 (2007).

\bibitem{add1} D. G. Ladeira and E. D. Leonel, Chaos {\bf 17}, 013119
(2007).

\bibitem{ref21} M. Feigenbaum, J. Stat. Phys. {\bf{21}}, 669 (1979).

\bibitem{ref22} J. P. Eckmann and D. Ruelle, Rev. Mod. Phys. {\bf{57}},
617 (1985).
\end{thebibliography}
\end{document}